# Analysis on reflection spectra in strained ZnO thin films[1]

T. Makino[A2], Y. Segawa[A], M. Kawasaki[B3] and A. Ohtomo[B]

[A]Photodynamics Research Center, RIKEN (The Institute of Physical and Chemical Research), Sendai 980-0845, Japan
[B]Institute for Materials Research, Tohoku University, Sendai 980-8577, Japan

Thin films of laser molecular-beam epitaxy grown ZnO films were studied with respect to their optical properties. 4-K reflectivity was used to analyze various samples grown at different biaxial in-plane strain. The spectra show two structures at ≈3.37 eV corresponding to the *A*-free exciton transition and at ≈3.38 eV corresponding to the *B*-free exciton transition. Theoretical reflectivity spectra were calculated using the spatial dispersion model. Thus, the transverse energies, the longitudinal transversal splitting ($E_{LT}$), the oscillator strengths, and the damping parameters were determined for both the *A*- and *B*-free excitons of ZnO. As a rough trend, the strain dependence of the energy $E_{LT}$ for the A-excitons is characterized by a negatively-peaking behavior with a minimum around the zero strain, while $E_{LT}$ for the B-excitons is an increasing function of the strain field values.

**Keywords**: Stresses, Laser epitaxy, Oxides, Semiconducting II-VI materials

**PACS**: 78.40.–q, 778.66.Li, 78.67.Pt, 81.15.Fg

Zinc oxide, a wurtzitic semiconductor, and its related oxides containing Mg or Cd have recently attracted a lot of attention due to potential applications for realizing short-wavelengths optoelectronic devices and surface acoustic wave modulators [1, 2]. Much effort has recently been devoted to the studies of ZnO epitaxy. This motivates also in-depth investigations of its optical properties and its electronic structures. ZnO epilayers oriented in the *c*-axis direction usually suffer from in-plane biaxial strain due to the mismatch of lattice constants and thermal expansion coefficients between the epilayer and the substrate. Quantum well (QW) structures also suffer from a certain amount of biaxial strain due to the lattice-mismatch between the well and barrier layers.

The in-plane lattice constant of sapphire(0001) substrate is largely mismatched with that of ZnO (18%). However, the actual amount of strain induced at the interface between ZnO and sapphire is likely to be deviated from the calculated value, and it depends on the epitaxy mechanism and hence growth conditions, *e.g.*, growth and annealing temperatures. It is important to quantitatively evaluate the amount of strain in actual ZnO epilayers and to achieve deeper understanding of its electronic structures under in-plain biaxial strain.

In a previous work [3], we reported the strain field dependence of the exciton resonance energies in strained and unstrained epilayers. The similar studies have been more recently conducted by other research groups [4, 5]. It is well known that an analysis of the reflection spectra can provide various information other than the exciton energies; oscillator strength, transverse-transverse splitting energies, and damping parameters. In this work, we determined these parameters by their line shape analysis, whose dependence on the strain field will be discussed.

The samples were undoped ZnO epilayers grown on sapphire substrates by laser molecular-beam epitaxy. We used various types of samples at various growth and annealing temperatures. Reflection spectra were taken at 4.2 K. The lattice constants were measured using the four-axis x-ray diffraction. The values for residual strain along the *a*- and *c*-axes were estimated by $\varepsilon_{xx}=(a-a_0)/a_0$ and $\varepsilon_{zz}=(c-c_0)/c_0$, where $a_0$ and $c_0$ are, respectively, the lattice constants in the unstrained crystal and taken as $a_0$=3.2505 Å and 5.2048 Å.

Figure 1 shows 4-K reflection spectra for seven different ZnO epilayers. Three transitions labeled "*A*", "*B*", and "*C*", which respectively refer to transitions from the corresponding valence bands to a conduction band. Three exciton series (*A*, *B* and *C*) are observed in wurtzite-type II-IV compounds. According to the polarization selection

---

[1] **conf52a249 (ICMAT2005)**
[2] electronic mail: makino@sci.u-hyogo.ac.jp, Present address: Department of Material Science, University of Hyogo, Kamigoori-cho, 678-1297, Japan
[3] Also at: Combinatorial Materials Exploration and Technology, Tsukuba 305-0044, Japan

rules, *A* and *B* transitions are allowed for light polarization perpendicular to the c-axis, while *C* transition is essentially allowed for polarization parallel to the *c*-axis [6, 7]. Because of the *c*-axis orientation, the *A* and *B* transitions are dominant. There is much information in the figure. First, we remark that it is very difficult to confirm the excitonic structure in the case of $\varepsilon_{zz}$=-0.11%, while one can observe easily distinct structures in the case of $\varepsilon_{zz}$=+0.30%. This asymmetric robustness in favor of the positive $\varepsilon_{zz}$ direction is interesting. Below the *A*-exciton resonance energy, some Fabry-Perot modes are observed, that are due to multireflection in the epilayer (curves for $\varepsilon_{zz}$=+0.065% and +0.30%). This is an indication of a good epitaxial relation between ZnO and $Al_2O_3$. Nevertheless, these Fabry-Perot (FP) modes do not overwhelm the oscillator structures observed around the exciton resonance regions. Therefore, for simplicity, we do not adopt a model so as to reproduce these FP modes in the spectral analysis. Let us pay attention to the peak-to-valley amplitudes (PVA) of the two "excitonic oscillators". For example, the PVA of the *A*-exciton relative to that for the *B*-exciton somewhat depends on the strain.

Here, we calculate the reflectivity by using the polariton model in order to estimate the strain dependence of exciton damping constants. In ZnO, there exists a strong coupling of free excitons with photons lying at similar energy, which is called the polaritonic effect. Calculation of polaritonic spectra in spatially dispersive media requires the so-called additional boundary condition. We use Pekar [8]'s model. The effective refractive index was used [9]. Contributions of the *C*-exciton and the excited states of the relevant three excitons are neglected. The dielectric function $\Xi$ is approximated by the following formula for the A-exciton[9];

$$\Xi(E,k)= \Xi_0(1+ E_{LT,(A)}/(E_{T,(A)}+\hbar^2 k^2/2M_{ex}-E-i\hbar\gamma_{(A)}/2 )). \quad (1)$$

Here, $\Xi_0$ is a background dielectric constant, $E_{LT,(A)}$, the longitudinal-transverse splitting energy, $M_{ex}$ is the effective mass of an exciton center-of-mass motion, and $\gamma_{(A)}$ is a damping constant. The parameter $\gamma$ has been replaced by a set of $\{\alpha_{(A)}, \beta_{(A)}\}$ given by a relation,

$$\hbar\gamma_{(A)} = \alpha_{(A)}+\Theta(E-E_{T(A)}) \times \beta_{(A)} \times (E-E_{T(A)})/E_{LT(A)} \quad (2),$$

because of inappropriateness of a constant damping parameter over the whole spectral range of interest. This relation is the same as that used by Hayashi for the analysis of $PbI_2$. Here, $\Theta$ stands for the unit step function. The following values were used [10]; $\Xi_0$=8.1 and $M_{ex}$=0.23$m_0$ where $m_0$ is the free electron mass in rest. Although it is well known that the effective masses of electrons and holes are strain dependent, for simplicity, the fixed value of the exciton mass is used. We also took an exciton-free layer, the so-called dead-layer effect into consideration [11, 12, 13]. The dead layer effects shift the phase of refractive index (the refractive index is a complex number). The non-zero value of the phase changes significantly the shape of reflection line.

Observed reflection spectra are given in Fig. 2(a)–(f) by dashed lines. These figures also illustrate the calculated reflectivities (by solid traces). The PVA and the energy distance between peak and valley positions for the two excitons are quite successfully reproduced in the calculated spectra.

Deduced sets of the running parameters (the resonance energies of *A*- and *B*-excitons $E_{T,(A)}$ and $E_{T,(B)}$, longitudinal transverse splitting energy of $E_{LT,(A)}$ and $E_{LT,(B)}$, and the damping constants of $\alpha_A$ and $\alpha_B$ are summarized in the Table I. The degrees of phase change $\theta$ were also plotted.

The exciton transition energies $E_{T,(A)}$ and $E_{T,(B)}$ are shown in Fig. 3 of Ref. [3]. When we plot the longitudinal-transverse splitting energy ($E_{LT,(A)}$, proportional to the oscillator strength) for the A-exciton state against $\varepsilon_{\{zz\}}$, we notice that this tendency can be fitted by a negatively peaking Gaussian function with a minimum of 1.6 meV around the zero strain, $\varepsilon_{\{zz\}}$=0. This seems to suggest that the mixing effect of valence bands is difference for the negative strain region and for the positive region. The strain dependence for $E_{LT,(B)}$ is, however, different for the abovementioned one. Namely, the maximum value of the longitudinal-transverse splitting energy of *B*-exciton (14.8 meV) is 31%-of-magnitude larger than the literature value ($E_{LT,(B)}$=11.1 meV), while the minimum value (11.0 meV) is very slightly smaller than the literature value It is very difficult to extract a certain tendency concerning the $E_{LT,(B)}$. This shows oscillating behavior with respect to the $\varepsilon_{zz}$. The obtained fitting parameters are randomly scattered. Reason of this complicated strain dependence of the *B*-exciton state has not been cleared yet.

The $\alpha_A$ and $\alpha_B$ locate within the range between 2.0 and 4.0 meV, and between 2.2 and 8.0 meV, respectively. It is realized that, by comparing the $\alpha$ between zero strain and $\varepsilon_{zz}$=+0.2956%, the quality of the strain-free epilayers is higher than that of the much strained sample.

In summary, low-temperature reflectivity was used to analyze various samples having biaxial in-plane strain $\varepsilon_{zz}$ induced by the difference between ZnO and (0001) sapphire substrates. Their spectra have mainly two structures corresponding to the *A*-free exciton and the *B*-free exciton transitions, respectively. Theoretical reflectivity spectra were calculated using the spatial dispersion model with two oscillators. Thus, the transverse energies, the longitudinal transversal splitting, the oscillator strengths, and the damping parameters were determined for both the *A*- and *B*-free excitons of ZnO. It is found that, by comparing the damping constant (α) between zero strain and $\varepsilon_{zz}$=+0.2956%, the quality of the strain-free epilayers is higher than that of the much strained sample.

The authors would like to acknowledge T. Yasuda for providing samples and experimental spectra. This work was funded in part by the MEXT Grant of Creative Scientific Research #14GS0204, the MEXT Grant-in-Aid for Young Scientists #15760015, the Asahi Glass Foundation, and the inter-university cooperative program of the IMR.

Table I: The parameters deduced by the fitting procedures in the six ZnO samples. Magnitude of the strain along *c*-axis of $\varepsilon_{zz}$, the resonance energies of *A*- and *B*-excitons $E_{T,(A)}$ and $E_{T,(B)}$, longitudinal transverse splitting energy of $E_{LT,(A)}$ and $E_{LT,(B)}$, and the damping constants of $\alpha_A$ and $\alpha_B$.

| $\varepsilon_{zz}$ (%) | $E_{T,(A)}$ (eV) | $E_{T,(B)}$ (eV) | $E_{LT,(A)}$ (meV) | $E_{LT,(B)}$ (meV) | $\alpha_A$ (meV) | $\alpha_B$ (meV) | $\theta$ (rad) |
|---|---|---|---|---|---|---|---|
| -0.075 | 3.374 | 3.383 | 3.8 | 11.3 | 3.3 | 5.5 | $3.1 \times 10^{-2}$ |
| -0.058 | 3.376 | 3.382 | 1.8 | 12.6 | 2.6 | 2.2 | $6.3 \times 10^{-2}$ |
| +0.023 | 3.378 | 3.385 | 1.4 | 13.8 | 3.5 | 5.5 | $6.3 \times 10^{-2}$ |
| +0.050 | 3.376 | 3.387 | 2.5 | 11.0 | 4.0 | 4.8 | $1.9 \times 10^{-1}$ |
| +0.065 | 3.375 | 3.381 | 4.5 | 14.8 | 2.0 | 2.8 | 6.3 x10⁻² |
| +0.300 | 3.377 | 3.384 | 5.6 | 12.0 | 3.9 | 8.0 | 0 |

Figure 1: Strain field dependence of reflectance spectra of ZnOsapphire epilayers measured at 4.2 K. Structures labeled "*A*", "*B*", and "*C*" are due to exciton resonances of *A*, *B*, and *C* transitions, respectively. Quantity of $\varepsilon_{zz}$ (%) shown on the right hand side represents the uniaxial strain along the *c*-axis, which is plus for tensile (biaxially compressive) and minus for compressive (biaxially tensile) strain.

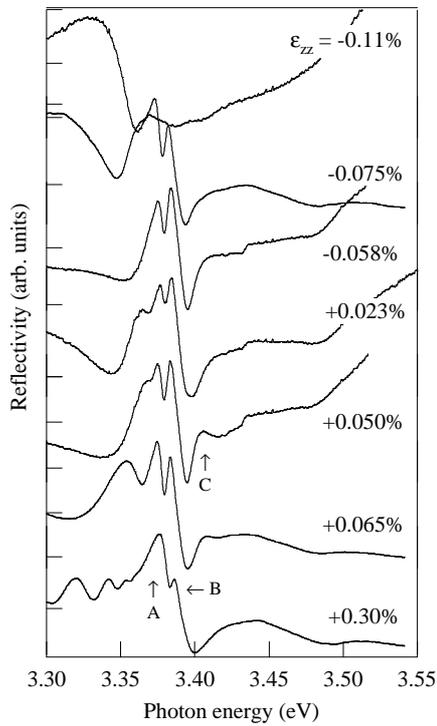

Figure 2: (a)–(f): Observed reflection spectra in ZnO epilayers (dashed lines), and calculated spectra (solid lines). Strains along the *c*-axis ($\varepsilon_{zz}$; %) were shown right-hand side. Constants used in each calculation are summarized in Table I.

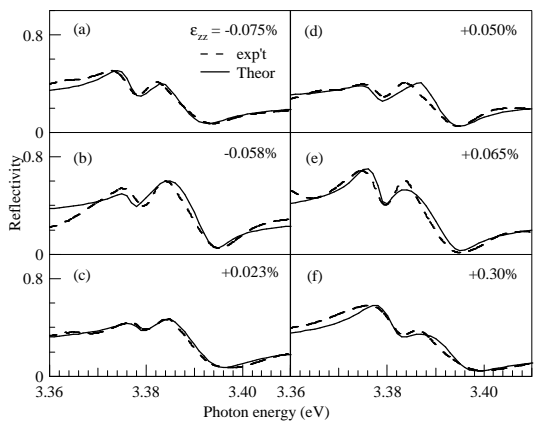